# Secondary Cosmic Muon Angular Distributions with High Statistics

J. Poirier, J. Gress, and T. F. Lin
*Physics Department, 225 NSH, University of Notre Dame, Notre Dame, IN 46556, USA*

## Abstract

Project GRAND presents a map in right ascension and declination for single cosmic secondary tracks which have been identified as muons. These muons are measured in stations of proportional wire chambers (PWCs) which have a mean angular resolution of 0.26°. A 50 mm steel plate interspersed with the PWCs is used to distingush muons from electrons. Two years of data are accumulated in a map of right ascension and declination which covers all of the northern hemisphere and a small portion of the southern hemisphere. The total number of muon tracks in this data sample are 40 billion. Significant large angular structure is found in this map; the deviations from average are small compared to the average, but because of the high statistics, the deviations from average are statistically significant to many standard deviations. Systematic effects in the data are minimized by averaging over this two year period. These results are similar to those obtained in previous data.

## 1 Introduction:

Project GRAND is a detector designed to measure the angles of secondary cosmic ray tracks and detect which tracks are muons. These muons come mostly from hadronic primary cosmic rays which interact with the earth's atmosphere producing pions which then can decay to muons in competition with their hadronic interactions. In addition to this muon production from hadronic primary cosmic rays, there is a 1.5% chance that a 100 GeV primary cosmic gamma ray would produce a single muon track at ground level (Fasso & Poirier, 1999). Regardless of their origin in the atmosphere, most of these secondary muons reach the earth where the surface detectors are located. A full-sky map of these secondary cosmic muons is useful in a search for possible point sources and to survey other interesting phenomena. A statistically significant large scale structure is observed. This paper updates previous work: (Fields et al., 1997; Poirier et al., 1995; Poirier et al., 1991); each of these references is a statistically independent data set. The deviations in right ascension are qualitatively similar to another experiment (Battistoni et al., 1988).

The probability that a primary cosmic gamma ray will yield a secondary muon at detection level is about 1.5%. This probability is calculated with FLUKA Monte Carlo code which predicts an increasing probability as the primary gamma ray energy increases (Fasso & Poirier, 1999). Thus the original idea in accumulating the data into these 5° x 5° pixels was the chance to observe point sources of primary gamma rays in the energy interval from 10 to 300 GeV. The problem with this approach is that there are 1584 such pixels, and the statistical significance of a positive signal in one such pixel has to be degraded by the fact that many pixels were examined. The overall chance of observing a statistically significant random point source of gammas with this technique is not overly encouraging. The chances of observing such a source(s) would be enhanced if the search was over a limited number of such possibilities with a known position in space (for D.C. sources) or in space and time (for periodic or burst sources). This approach is used in the following papers to this conference: (Carpenter et al, 1999; Roesch et al, 1999).

During this exercise, a large scale structure was noted which is very significant for the following reasons: 1) statistically, the number of sigma is very large, 2) the angular scale of the excess or deficiency is so large that few could be in the entire interval searched, thus limiting the number of possibilities, 3) the data are averaged over a two year period, and 4) similar results have been obtained in prior year's data which are statistically independent. The details of the analysis are described below.

## 2  Experimental Array:

The extensive air shower array is located at 220 m above sea level at 86° W and 42° N. It is composed of 64 stations of position sensitive detectors [proportional wire chambers (PWCs)] each with an active area of 1.25 m$^2$. Each station has four such PWCs arranged above each other with a space in between so as to geometrically measure the angles of secondary cosmic ray tracks. Each of these detectors is composed of two orthogonal proportional planes which allow measurement of orthogonal positions in the $X$ (East) and $Y$ (North) directions. These position measurements allow a determination of the angles in the $XZ$ and $YZ$ directions ($Z$ is the vertical direction). There is a 50 mm steel plate positioned immediately above the bottom (fourth) PWC which allows differentiation between muon and non-muon tracks; that is, muons pass through the steel without stopping or interacting 96% of the time, whereas non-muon tracks fail to do this 96% of the time. The present rate of accumulating identified muon angle data is 170 million per day. Accumulating data of single track muons is concurrent with running on multiple station triggers which is rich in extensive air shower data; the rate of shower data is 1.4 per sec or about a million per week.

## 3  Data Analysis:

Project GRAND's average angular resolution for identified secondary muons at detection level is 0.26° in the $XZ$ and $YZ$ planes. However, the angular resolution for the primary cosmic ray which produced the muon is degraded by the fact that: the pions have a production angle relative to the primary; the muons have a decay angle relative to the pion; and the muons scatter in the remaining air thickness and bend in the earth's magnetic field. The combination of these effects degrades the angular resolution for the primary to about ± 5°. The data were therefore combined into cells of 5° x 5° pixels.

Data are selected only from those runs that contain an integer number of sidereal days for these measurements; 300 days of complete sidereal data were recorded. The right ascension (RA: $\alpha$) and declination (Dec: $\delta$) are then calculated for the time of that data. The data are then examined for uniformity and continuity. Each sidereal day's data has been checked for smoothness before inclusion in the summed data. The smoothness test was conducted in the following manner: The numbers in the 110° of $\delta$ were projected onto the $\alpha$ axis for each sidereal day. For each day, the difference between the maximum and minimum number is divided by the average. For a sidereal day's data to be included in the final sum, their worst variation, either maximum or minimum, were required to be less than ± 5% from their average number. There were 254 sidereal day's data which passed this smoothness test; these smooth data were accumulated into one file which summed the data in a two dimensional grid of 1° x 1° intervals in $\alpha$ from 0° to 360°, and in $\delta$ from -20° to 90°. The number of smooth sidereal days in each month from Jan. to Dec. for these two years is as follows: 29, 27, 9, 21, 12, 15, 18, 11, 23, 24, 30, 35, giving a minimum of 9 in March and a maximum of 35 in December. The raw data thus consist of this array $N(\alpha,\delta)$ of accumulated counts. The total number of single track muons which were accumulated in this fashion totaled 40 billion.

The variation of the muon angles in declination (having summed over $\alpha$, $<N(\delta)>$) is shown in Figure 1. $<N(\delta)>$ is a strong function of $\delta$ because near zenith is where GRAND's detection efficiency and the average secondary cosmic ray rate both are at a maximum. The variation of single muon data with $\alpha$ is shown in Figure 2. In contrast with Figure 1, these data are extremely flat (notice surpressed zero); the variation here is ± 0.3% from average. It is convenient to take into account the strong variations of muon detection over declination angles to plot muon angles in $\alpha$- and $\delta$-space. To do this, bins are added and combined into 5° x 5° pixels. At each declination, the difference $N(\alpha,\delta) - <N(\delta)>$ is calculated (where $<N(\delta)>$ is $N(\alpha,\delta)$ averaged over all $\alpha$). This difference of each $\alpha,\delta$-pixel from its average is then normalized by dividing by its standard deviation (s.d.):

$$\text{diff}(\alpha,\delta) = (N(\alpha,\delta) - <N(\delta)>) / \text{s.d.} \qquad \text{where} \qquad \text{s.d.} = \sqrt{<N(\delta)>}]$$

The results of the deviations of each 5° x 5° pixel in units of its standard deviation are shown in a contour plot, Figure 3. This plot shows lines which connect those regions with the same number of standard deviation's difference($\sigma$) from the average for that declination. A large-scale region of excess (more than +23$\sigma$) is centered near $\alpha = 320°$ and $\delta = 40°$ and a large-scale region of deficiency (more than -25$\sigma$) located near $\alpha = 150°$ and $\delta = 40°$. As can be seen from the numbers, these differences are statistically very significant--the further so when it is noted that these deviations are for a single 5° x 5° pixel. The number of such pixels contained inside the -25$\sigma$ contour is 36 and inside the +23$\sigma$ contour is 50.

Of course, the concern of such small statistical errors is the possible systematic effects which could cause these deviations (the errors quoted are statistical only). To check one such theory: In the summer the weather is good and with extra full-time student help, it is a traditional time to do a lot of work in the field. This work sometimes entails turning off a single hut for a time. Since the rest of the field can continue to take data, the run is usually continued while the one hut is turned off. This would cause a systematic deficiency at those RA bins which are near zenith for the workday hours of those summer months. A small change like this could escape the smoothness test for that day and thus accumulate as many day's data are added together. However, the mid-day zenith during the summer months is $\alpha$ near zero, a region near maximum deviation (rather than the expected minimum predicted for this systematic effect).

## 4  Conclusions:

It is well known that gamma rays at EGRET energies are enhanced toward the galactic disk which has somewhat the pattern as shown in Fig. 3. While the angular scales suggest the galactic disk, the maximum and minimum in the plot do not, however, coincide with the angular position of the disk; indeed, the disk partially overlaps both the enhancement and the deficit. Thus, the contribution of the disk (and disk gamma rays) to the muon angular distribution remains unclear. Local cosmic ray anisotropies could have a role; indeed, the fluctuations are of the order of cosmic ray anisotropies like the Compton-Getting Effect reported in the literature (Culter & Groom, 1991). Local effects could also come into play; e.g., variations in the solar wind magnetic field could perhaps account for these features. An air-tide would correlate the data with *solar* time. A uniform exposure over the solar year would average this effect so as to not appear in a plot of *sidereal* angles. Further work on these issues is clearly in order.

As for smaller angular-scale sources, the evidence is not strong. Within the experimental sensitivity and resolution, there is no obvious signal of individual point sources above the normal expected statistical fluctuations in the 1584  5° x 5° cells that are searched.

This research is presently being funded through grants from the University of Notre Dame and private donations. The National Science Foundation participated in GRAND's construction.

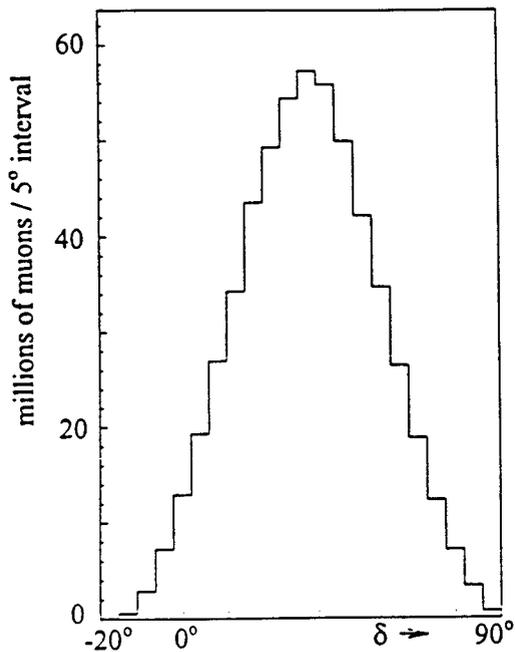

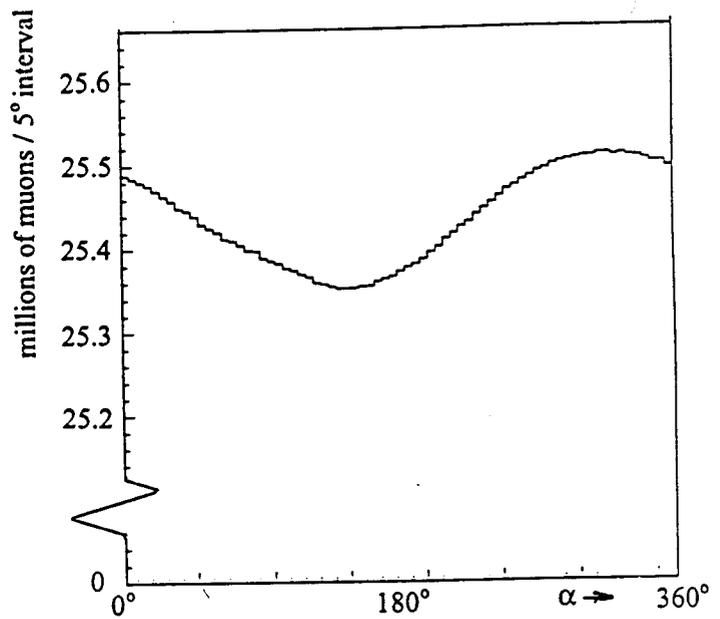

**Figure 1:** Histogram of muons in each 5° declination interval projected from 360° in right ascension.

**Figure 2:** Histogram of muons in each 5° right ascension interval projected from 110° in declination. The curve is flat to ± 0.4% (note surpressed zero).

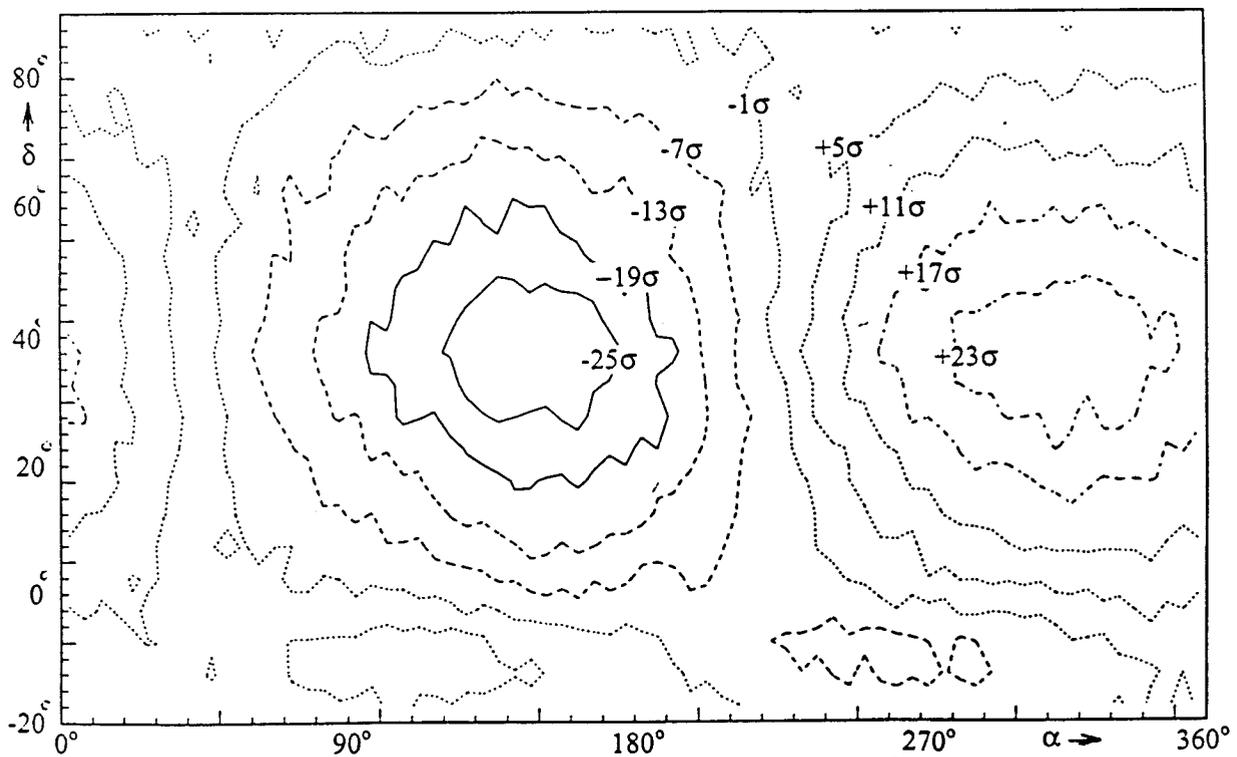

**Figure 3:** Contour map of muon rate deviations as a function of $\alpha$ and $\delta$; the deviations are in units of $\sigma$. The numbers indicate the extent of the deviation of a 5° x 5° pixel from its average.